# Ultrafast Manipulation of Valley Pseudospin

Authors: Ziliang Ye[1,2] *, Dezheng Sun[1,2]*, Tony F. Heinz[1,2%]

[1] *Departments of Applied Physics and Photon Science, Stanford University, 348 Via Pueblo Mall, Stanford, CA 94305, USA*

[2] *SLAC National Accelerator Laboratory, 2575 Sand Hill Rd., Menlo Park, CA 94025, USA*

**The coherent manipulation of spin and pseudospin underlies existing and emerging quantum technologies, including NMR, quantum communication, and quantum computation[1,2]. Valley polarization, associated with the occupancy of degenerate, but quantum mechanically distinct valleys in momentum space, closely resembles spin polarization and has been proposed as a pseudospin carrier for the future quantum electronics[3,4]. Valley exciton polarization has been created in the transition metal dichalcogenide (TMDC) monolayers using excitation by circularly polarized light and has been detected both optically[5-7] and electrically[8]. In addition, the existence of coherence in the valley pseudospin has been identified experimentally[9]. The *manipulation* of such valley coherence has, however, remained out of reach. Here we demonstrate an all-optical control of the valley coherence by means of the pseudomagnetic field associated with the optical Stark effect. Using below-bandgap circularly polarized light, we experimentally rotate the valley exciton pseudospin in monolayer WSe$_2$ on the femtosecond time scale. Both the direction and speed of the rotation can be optically manipulated by tuning the dynamic phase of excitons in opposite valleys. This study completes the generation-manipulation-detection paradigm for valley pseudospin, enabling the platform of excitons in 2D materials for the control of this novel degree of freedom in solids.**

The broken inversion symmetry in monolayer TMDC crystals gives rise to a nontrivial Berry phase at the K and K' valleys in momentum space[10], where the optically accessible direct band gap occurs in these materials[11,12]. This Berry phase, together with the angular momentum of the atomic orbitals, leads to different optical selection rules for the two valleys: Excitons in the K valley couple to left-circularly polarized photons, while excitons in the K' valley couple to right-circularly polarized photons, thus permitting the optical generation of valley polarization through control of the helicity of light[5-7]. Moreover, when a TMDC monolayer is excited with linearly polarized light, a coherent superposition state is established between the K and K' valley excitons[9]. Such a coherent state can, in principle, be manipulated by lifting the energy degeneracy between valleys through the breaking of time-reversal symmetry in the material. In this regard, dc magnetic fields have been applied to achieve a valley splitting of a few meV[13-16]. Circularly polarized light also breaks time reversal symmetry, and researchers have used it to create larger valley splittings, corresponding to pseudomagnetic fields up to 60 T[17,18]. In addition, this optical approach permits us to access the ultrafast time scale: In this work, we demonstrate the coherent manipulation of the valley pseudospin on the femtosecond time scale.

The optical Stark effect has been applied to manipulate the spin/pseudospin in quantum coherent systems, such as atomic quantum gases[19], quantum dots[20,21], and III-V quantum wells[22]. Here we utilize this approach of a strong pseudomagnetic field in monolayer $WSe_2$ to rotate the valley pseudospin in the superposition plane of K and K' valley polarizations (Fig. 1). We start by exciting the monolayer with quasiresonant linearly polarized light to generate a coherent superposition state. In the Bloch sphere representation, this is equivalent to a pseudospin lying in the equatorial plane, with the |K> and |K'> states pointing towards the north and south poles, respectively. Following this coherent excitation, we apply a below-gap circularly polarized control pulse to lift the valley energy degeneracy by $\hbar\Delta\omega$. The induced change in the transition energy of the two valleys leads to different oscillation frequencies for excitons at K and K' valleys and, accordingly, to the development of a dynamic phase difference $\Delta\phi \sim \Delta\omega \cdot \Delta t$ for a control pulse of duration $\Delta t$. We can describe the effect of this phase evolution as a rotation of the valley pseudospin in the equatorial plane of the Bloch sphere by an angle $\Delta\phi$.

We first use transient reflectivity spectroscopy to characterize the breaking of the valley degeneracy by the optical Stark effect, as previously explored in Ref. 17 and 18. (See Methods for details concerning our measurements.) As shown in Fig. 1c, the exciton peak in monolayer $WSe_2$ is blue shifted by about $2\pm0.5$ meV for a circularly polarized control pulse with a fluence of 1 mJ/cm$^2$ (at the sample after correcting for the influence of the substrate) with a red detuning of 160 meV from the A exciton. This optical Stark shift, comparable to that reported previously[17,18], occurs only in one valley and induces an energy splitting between the K and K' valleys. (The peak shift observed in a transient reflectivity experiment is somewhat smaller than the intrinsic Stark shift of 10 meV that would be achieved for a longer control pulse of the same intensity. This situation arises because the duration of control pulse is comparable to the total coherence time of the excitonic transition, including both homogeneous and inhomogeneous contributions, so that averaging over the time-dependent control pulse leads to a reduced shift. We analyze this effect in detail in the Supplementary Information, Sect. IV.) The energy splitting implies a pseudospin beat frequency of $\Delta\omega/2\pi = 0.48$ THz. Based on our control pulse duration of $\Delta t \sim 100$ fs, we expect an induced pseudospin rotation of roughly $0.1\pi$, in qualitative agreement with the experimental results presented below.

We tracked the valley pseudospin in the Bloch sphere (Fig. 1d) through its signature in the polarization of the photoluminescence[9]. Following application of the linearly polarized excitation pulse and of the circularly polarized control pulse, we collect the PL signal and analyze its polarization state (Methods). Since the linearly polarized excitation generates valley coherence between excitons in the TMDC monolayer, the PL exhibits a linear polarized component with its axis aligned with the excitation direction. (In the ideal case, the emission would be fully linearly polarized for linearly polarized excitation of the two degenerate valleys; dephasing effects reduce the degree of linear polarization, but do not change the orientation.) If the valley exciton pseudospin is rotated in the Bloch sphere by angle $\Delta\phi$, the axis of the PL polarization will rotate by half the corresponding angle $\Delta\phi/2$ (Supplementary Information, Sect. I).

Thus, by examining the polarization of the PL after application of our control pulse, we can measure the direction and speed of the pseudospin rotation.

Experimentally, we first measure the influence of the control pulse by examining the polarized PL using an optical bridge scheme. For linearly polarized excitation along the **x** direction (Fig. 2), we set the analyzer to $\theta = 45°$ and collect the parallel (**a** channel) and perpendicular (**b** channel) polarization components of the photoluminescence. The A exciton PL peak in WSe$_2$ is observed at 1.725 eV, with a Stokes shift of about 20 meV and a small trion peak at 1.701 eV, sitting on a background from localized states and higher-lying transitions. In the absence of the control pulse, the polarization axis of the emission coincides with the excitation direction: the two detection channels then yield identical spectra (Fig 2, open dots). For a left circularly polarized control pulse introduced nearly simultaneously with the excitation pulse, the response in the two channels (Fig. 2, solid dots) differs near the A peak, with the **a** channel increasing significantly for photon energies from 1.712 to 1.738 eV. This indicates a counter-clockwise rotation of the valley pseudospin induced by the control pulse.

We summarize the influence of the polarization rotation by recording the PL intensities $I_a$ and $I_b$ of the A exciton in the two orthogonal detection channels. (The A exciton strength is obtained by fitting the experimental emission spectra and excluding the background contribution.) The degree of polarization is described by the normalized Stokes parameter $S_2 = (I_a - I_b)/(I_a + I_b)$ with our analyzer set at $\theta = 45°$. As shown in the inset of Fig. 2, $S_2$ increases monotonically with the power of the control beam. This trend is expected provided that the induced rotation of the polarization doesn't exceed *π/4* for the highest power in our experiment.

To measure the orientation of the polarization of the PL more precisely, we have performed measurements in which we scan the analyzer angle $\theta$ (Fig. 3, with complete spectra in Supplementary Information, Sect. V). For arbitrarily polarized light with a linear polarization component, $S_2(\theta)$ varies as $cos[2(\theta - \theta_0)]$, where $\theta_0$ is the angle of the linearly polarized component of the radiation with respect to the **x**- direction. In Fig. 3, we show the results for different polarization states of the control pulses. When the control pulse is linearly polarized, $S_2$ follows the expected sinusoidal variation with $\theta_0 = 0$, corresponding to no rotation of the valley pseudospin. However, for a left circularly polarized control pulse (blue curve), the angular dependence is clearly shifted, corresponding to $\theta_0 = 0.12\pi = 22°$ or a 44° counter-clockwise rotation of the valley pseudospin in the Bloch sphere. For a right circularly polarized control pulse (red curve), the angular dependence is shifted in the opposite direction by $\theta_0 = -0.11\pi$. This result directly confirms that the direction of polarization rotation of the PL (and, hence, of the underlying valley pseudospin) is determined by the helicity of the circularly polarized control pulse. We note that the amount of pseudospin rotation, while constrained here by our detection scheme, should readily cover the full $2\pi$ range. (The maximal rotation is limited by the effect of the two-photon PL induced by the control pulse, which interferes with the one-photon PL generated by the excitation pulse. In the future, the two-photon PL can be avoided by, for example, limiting the control photon energy to be less than half of the exciton transition energy.)

In addition to the rotation of the PL polarization, we also observe a reduction in the maximum value of $S_2$. In contrast, when the control pulse either precedes the excitation pulse or is linearly polarized, we find change in neither the orientation of the polarization nor in the magnitude of $S_2$. The depolarization effect observed under pseudospin rotation can be understood as the consequence of the incoherent integration of the PL in both the temporal and spatial domains (Supplementary Information, Sect. II), and no additional pump-induced decoherence processes need to be invoked.

By using accurate control of the delay time of the control pulse after the excitation pulse, we can directly probe the dynamics of the valley coherence (Fig. 4). For each choice of excitation-control delay, we characterize the PL by the value of $S_2$ ($\theta$ = 45°). We observe that $S_2$ peaks for a time delay of 50 fs, which results from the maximal pseudospin rotation in Fig. 3, and vanishes after 800 fs. This fast response suggests that the intervalley decoherence time $T_2$ in monolayer WSe$_2$ is also in the range of hundreds of femtoseconds. If we switch the helicity of the control pulse, the delay dependence of $S_2$ is inverted, in agreement with the rotation picture described above.

To analyze the ultrafast dynamics of valley coherence, we introduce a quantum beat model of the process. For simplicity, we assume instantaneous excitation of the superposition state along the **x** direction at time t = 0. We then find that the PL emission along the $\theta$ direction is proportional to $I_\theta(t) \propto e^{-t/T_1}(\rho_{11}^0 + \rho_{22}^0) + 2e^{-t/T_2}\rho_{12}^0 Cos(\phi_1(t) - \phi_2(t) - 2\theta)$ (Supplementary Information, Sect. I). Here $\rho_{11}$ and $\rho_{22}$ are the diagonal elements of the density matrix associated, respectively, with excited states at K, K' valleys and the ground state, and $\rho_{12}$ is the corresponding off-diagonal element. $T_1$ is the exciton population lifetime; $T_2$, the intervalley decoherence time; and $\phi_i(t)$ (for $i = 1,2)$, the dynamic phase, proportional to the time integral of the time-varying excited-state energy $E_i(t)$. Since we do not time resolve the PL in our experiments, but rather integrate the signal over time, we compute the delay dependence of the Stokes parameter $S_2$ accordingly (Fig. 4, solid line). To apply this model, we use the known duration of the control pulse and the fact that that maximum value of $S_2$ is 20% in the absence of a control pulse. This information can be used to estimate the ratio between $T_1$ and $T_2$, leaving us with only two independent parameters (Supplementary Information, Sect. I). (We note that the partial degree of linear polarization has several possible origins, including depolarization induced by defects/impurities and the exchange interaction between excitons.[23-25]. Within our model, we neglect possible initial depolarization effects in creating the valley excitation, in view of the near-resonant character of the optical excitation.) We are then left with only two independent parameters: The intervalley decoherence time $T_2$, and the optical Stark shift. We obtain a match to experiment for $T_2 = (350 \pm 50)$ fs, a value comparable to the intravalley decoherence time measured by a recent photon-echo experiment[26]. Moreover, the fitted Stark shift also agrees with value determined by our frequency-domain measurements (Supplementary Information, Sects. III and IV).

In conclusion, we have demonstrated the possibility of ultrafast rotation of the exciton valley pseudospin in monolayer WSe$_2$, with controlled rate and rotation angle. Our scheme for valley pseudospin manipulation can be readily extended to the recently discovered quantum-confined states of excitons in TMDC layers, the lifetime of which is orders of magnitude longer than our free excitons[27-30]. Due to the strong spin-valley coupling in TMDCs, valley pseudospin is expected to be well preserved in these localized states[31]. In addition, full control of valley pseudospin could be achieved by means of stimulated Raman adiabatic passage in a real or pseudo magnetic field. More generally, excitons in TMDC layers play a central role in emerging optoelectronic applications of 2D van-der-Waals materials and their heterostructures. The ability to manipulate valley coherence provides access to new degree of freedom for future quantum devices.

**Methods**

**Sample preparation and transient reflectivity measurements**

We mechanically exfoliated monolayer WSe$_2$ samples from bulk crystals onto silicon substrates covered by 270-nm SiO$_2$ layer. The sample dimensions exceed 5x5 µm$^2$, permitting the excitation and control beams to be fully confined to the monolayer. The samples were characterized by both Raman and PL spectroscopy. All experiments were performed at a temperature of 20 K.

We determined the optical Stark shift induced by the control pulse in the frequency domain by means of transient reflectivity measurements[17,18]. The spectra were obtained in a pump-probe configuration using a reflection geometry and zero time delay. The photon energy of the control (pump) pulse here and in the following experiments was chosen to be 1.57 eV, approximately 160 meV below the A exciton transition. The induced response was probed by an ultrafast supercontinuum pulse, covering the photon energy range from 1.71 eV to 1.79 eV. The required pulses were derived from the second-harmonic of an amplified mode-locked fiber laser (Impulse,

Clarke-MXR) operating at a wavelength of 1.03 $\mu$m. The 1.57 eV pulses were generated by a non-collinear optical parametric amplifier (NOPA), while the continuum pulses were produced by focusing fiber laser pulses of about 1 µJ energy in an undoped YAG crystal. Both pulses have a 100-fs full width at half maximum, calibrated by the rising part of the transient reflection signal from a gold thin film. The repetition rate for these measurements and for the pseudospin rotation experiments was 1 MHz. The pump and probe beams were focused at the sample with respective diameters of 2.1 µm and 1.7 µm. The finite size and temporal duration of the pump pulse reduced the optical Stark shift measured by the probe pulse. (The corresponding corrections are analyzed in Supplementary Information, Sect IV.) The pump beam was chopped at 700 Hz and the resultant modulation of the probe beam was measured by a lock-in technique. The measured reflection contrast spectra with and without pump excitation are presented in the Supplementary Information (Sect. III).

**Polarization-resolved photoluminescence measurements**

Measurements of the valley pseduospin rotation were performed by the study of polarization-resolved photoluminescence in the presence of sub-gap control pulses. The control pulses at a photon energy of 1.57 eV were obtained from a NOPA as described above and the desired polarization state was defined by a Babinet-Soleil compensator. Excitation pulses were obtained from filtered supercontinuum radiation with a photon energy of 1.81 eV, 85 meV blue detuned from the A exciton resonance, in order to maximize the linearly polarized coherent emission.[32] The fluence of the excitation pulses was limited to 1 µJ/cm² to avoid saturation and the fluence of the control pulses did not exceed 1 mJ/cm². The photoluminescence was recorded over a spectral range defined by 700-nm long-pass and 750-nm short-pass filters. The polarization characteristics of the PL were analyzed by rotation of a half-wave plate followed by a polarizing beam splitter. Measurement of both of the resulting signals (**a** and **b** channels) yielded significant reduction in common-mode noise in the system.

**Acknowledgment:** This work was supported by the Department of Energy, Office of Science, Basic Energy Sciences, Materials Sciences and Engineering Division, under Contract DE-AC02-76SF00515 and by the W. M. Keck Foundation.*These authors contributed equally to the work

%Corresponding author: tony.heinz@stanford.edu9

**Figure and Figure Captions:**

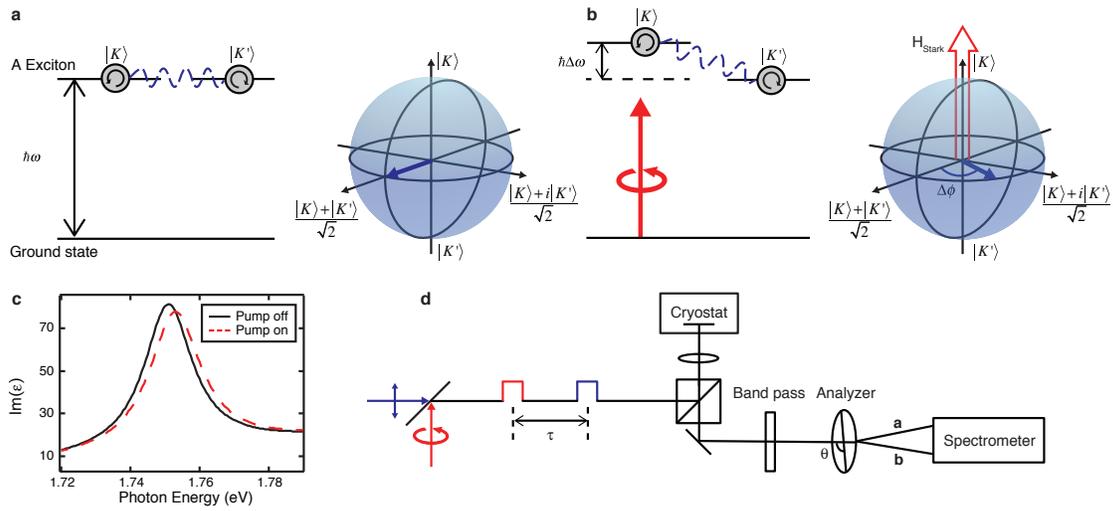

**Fig. 1** Schematic representation of all-optical valley pseudospin manipulation. a) A linearly polarized optical pulse excites a coherent superposition of K and K' excitons at optical band-gap energy $\hbar\omega$. The two equal components in the K and K' valleys have a fixed phase relationship, as illustrated by the wavy line and by the pseudospin lying in the equatorial plane of the Bloch sphere (right). b) When a strong left circularly polarized control pulse of below-gap radiation is applied (red arrow), the exciton transition energy in the K valley is increased by $\hbar\Delta\omega$. While the control pulse is present, a dynamic phase difference $\Delta\phi$ develops between the exciton components in the two valleys. The pseudospin in the Bloch sphere is correspondingly rotated by an angle $\Delta\phi$, just as would occur for an out-of-plane pseudomagnetic field (red open arrow). c) Results of a transient reflectivity measurement, indicating a valley energy splitting of $2\pm0.5$ meV produced by the optical Stark effect with a circularly polarized control pulse. The black solid and red dashed peaks are, respectively, the imaginary part of the fitted dielectric functions with and without the pump pulse. d) Experimental configuration used to generate and monitor the rotation of the valley pseudospin. An ultrafast control pulse is introduced at a controlled time delay $\tau$ after the linearly polarized excitation pulse. The photoluminescence generated by the excitation pulse is spectrally filtered, analyzed in terms of its polarization state, and collected by an imaging spectrometer.

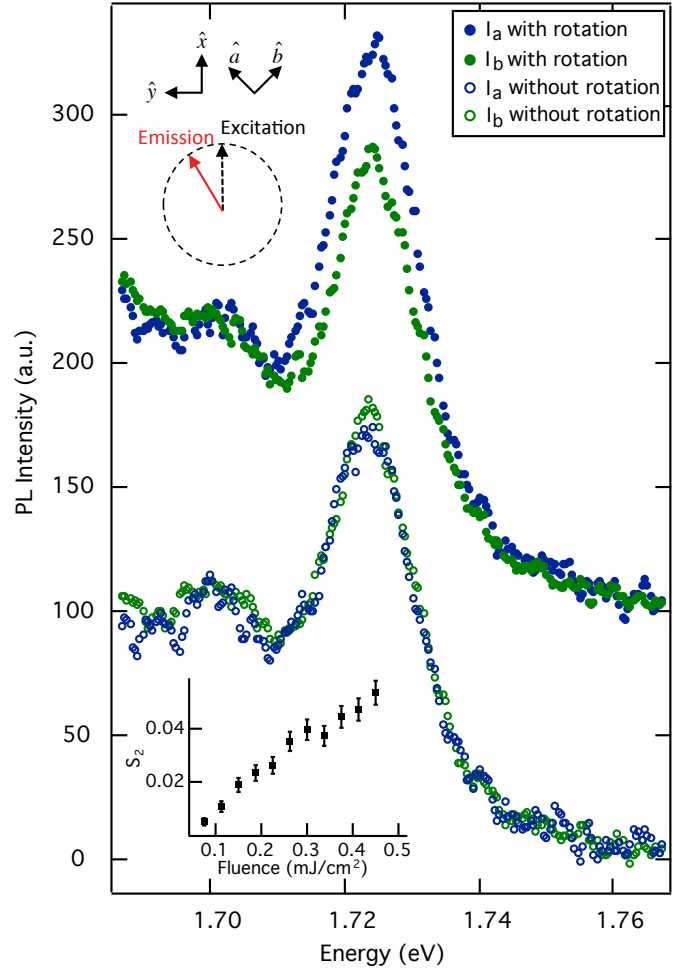

**Fig. 2** Photoluminescence spectra for both the rotated (solid dots) and unrotated (open dots) valley pseudospin. The emission both parallel (**a** channel, blue dots) and perpendicular (**b** channel, green dots) to the analyzer is collected. Here the analyzer is set at an angle of $\theta = 45°$ with respect to the linearly polarized excitation. For a slight time delay of the control pulse with respect to the excitation pulse ($\tau = 50$ fs, solid dots), the PL of the **a** channel is clearly stronger than that of the **b** channel PL, indicating rotation of the valley pseudospin. For the control pulse preceding the excitation pulse ($\tau$ = -200 fs, open dots), no rotation is observed under the same conditions. The PL peaks at 1.701 eV arise from trion states, which do not exhibit intervalley phase coherence and are unaffected by the control pulse. The inset shows the monotonic increase of $S_2$ with fluence of control pulse, indicating the presence of an induced rotation of less than 45° under the experimental conditions.

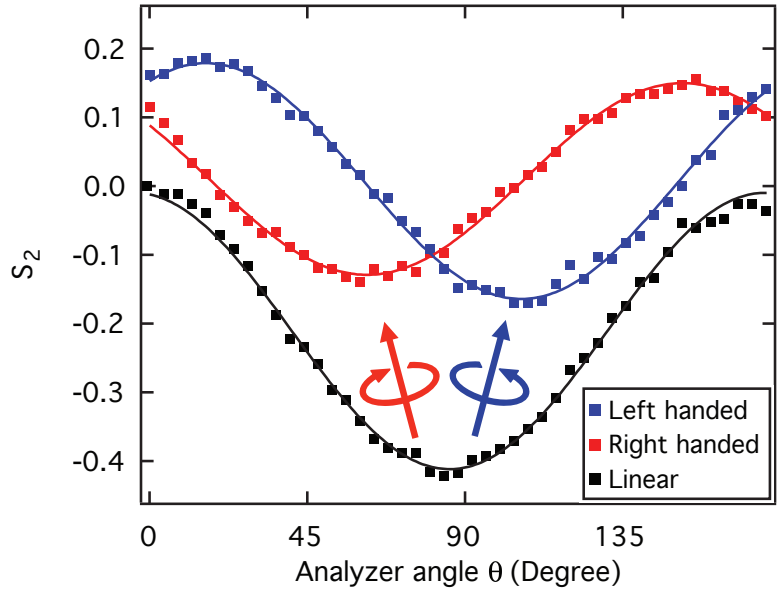

**Fig. 3** The rotation of the valley pseudospin is determined by measuring the orientation of the polarization of the photoluminescence. As the analyzer setting $\theta$ is scanned over 180°, we collect both **a** (parallel) and **b** (perpendicular) channels to determine $S_2$ as a function of $\theta$. When the control pulse is linearly polarized (black dots), $S_2$ follows $cos\, 2(\theta - \theta_0)$ with $\theta_0 = 0$ (black line), *i.e.*, without any discernible angular shift. For a right circularly polarized control beam, the emission shifts to $\theta_0 = -0.11\pi \approx -20°$ (red dots), corresponding to a clockwise rotation of the pseudospin by 40°. For a left circularly polarized control beam, the polarization shifts in the opposite direction to $\theta_0 = 0.12\pi$ (blue dots). Lines are the sinusoidal fits. For clarity, the values of $S_2$ for the linearly polarized control beam are offset by $-0.2$.

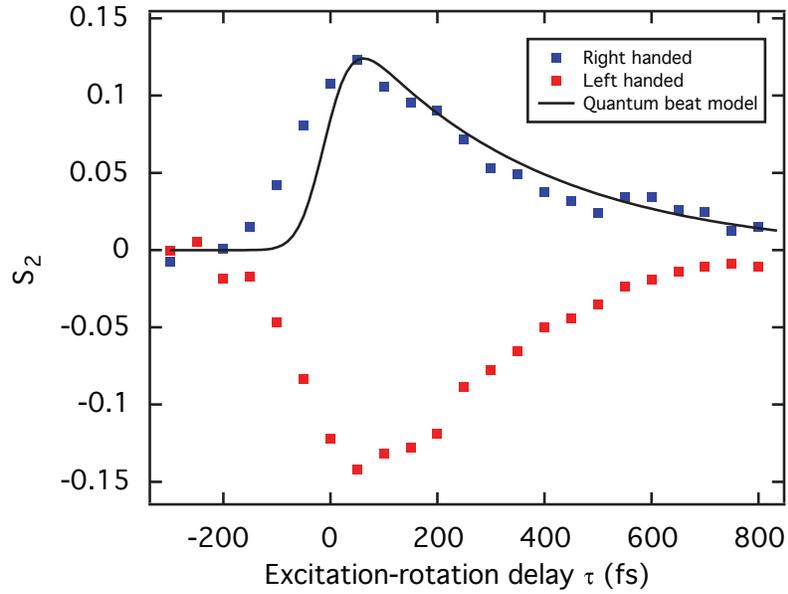

**Fig. 4** Determination of the intervalley decoherence time $T_2$ from the measurement of the dependence of $S_2$ on the excitation-control delay time. For a left circularly polarized control pulse, the pseudospin is rotated counter-clockwise, leading to an increased value of $S_2$ (blue dots). The response peaks for delay of $\tau \sim 50$ fs and disappears after $\tau \sim 800$ fs. The experimental result can be fit to the predictions of a simple quantum beat model (black line), yielding an intervalley decoherence time $T_2 = 350$ fs. For the opposite circularly polarized state of the control pulse, the response is inverted (red dots), as expected for a clockwise rotation of the valley pseudospin.